\documentclass[acmsmall]{acmart}
\AtBeginDocument{%
  \providecommand\BibTeX{{%
    \normalfont B\kern-0.5em{\scshape i\kern-0.25em b}\kern-0.8em\TeX}}}

\setcopyright{acmcopyright}
\copyrightyear{2023}
\acmYear{2023}
\acmDOI{XXXXXXX.XXXXXXX}

\acmJournal{JACM}
\acmVolume{37}
\acmNumber{4}
\acmArticle{111}
\acmMonth{8}

\usepackage{svg}

\usepackage{multirow}

\usepackage{xcolor}
\usepackage{ifthen}
\newboolean{showcomments}
\setboolean{showcomments}{true} 
\ifthenelse{\boolean{showcomments}}
    {
        \newcommand{\song}[1]{\textcolor{gray}{{\it [Song says: #1]}}}
        \newcommand{\emelie}[1]{\textcolor{purple}{{\it [Emelie says: #1]}}}\newcommand{\per}[1]{\textcolor{blue}{{\it [Per says: #1]}}}
    }
    {
        \newcommand{\song}[1]{}
       \newcommand{\emelie}[1]{} \newcommand{\per}[1]{}
    }

\setcopyright{none}
\renewcommand\footnotetextcopyrightpermission[1]{}
\begin{document}

\title{Industry Practices for Challenging Autonomous Driving Systems with Critical Scenarios}

\author{Qunying Song}
\authornote{Corresponding author.}
\email{qunying.song@cs.lth.com}
\orcid{0000-0002-8653-0250}
\authornotemark[1]
\author{Emelie Engström}
\email{emelie.engstrom@cs.lth.com}
\author{Per Runeson}
\email{per.runeson@cs.lth.com}
\affiliation{%
  \institution{Department of Computer Science, Lund University}
  \streetaddress{Box 118}
  \city{Lund}
  \country{Sweden}
  \postcode{SE-221 00}
}

\renewcommand{\shortauthors}{Song et al.}

\begin{abstract}
Testing autonomous driving systems for safety and reliability is extremely complex. A primary challenge is identifying the relevant test scenarios, especially the critical ones that may expose hazards or risks of harm to autonomous vehicles and other road users. 
There are several proposed methods and tools for critical scenario identification, while the industry practices, such as the selection, implementation, and limitations of the approaches, are not well understood. In this study, we conducted 10 interviews with 13 interviewees from 7 companies in autonomous driving in Sweden. We used thematic modeling to analyse and synthesize the interview data. We found 
there are little joint efforts in the industry to explore different approaches and tools, and every approach has its own limitations and weaknesses. To that end, we recommend combining different approaches available, collaborating among different stakeholders, and continuously learning the field of critical scenario identification and testing. The contributions of our study are the exploration and synthesis of the industry practices and related challenges for critical scenario identification and testing, and the potential increase of the industry relevance for future studies in related topics.
\end{abstract}




\maketitle

\fancyfoot{}
\thispagestyle{empty}


\section{Introduction}
\label{sec:introduction}
Before autonomous driving systems can be deployed on a large scale, they need to be rigorously tested to ensure that they can operate safely and reliably in a wide range of real-world scenarios. This testing process is complex and involves a variety of approaches, from virtual simulations to physical testing on closed tracks and public roads~\cite{lou2022testing}. In this paper, we explore the practical aspects of how autonomous driving systems are tested, with a special focus on the identification and use of critical scenarios for testing. The complexity and uncertainty of the driving environments and tasks lead to a potentially infinite number of test scenarios for autonomous driving systems, including many critical scenarios that may cause risks of harm to the autonomous vehicle and other road users~\cite{zhang2022finding, cai2022survey}. 

As articulated by the SOTIF (Safety of the Intended Functionality) standard, all relevant scenarios for autonomous vehicles should be tested, especially the critical ones, e.g., inclement weather, bad road conditions, and invasive behaviors from other road users, to ensure their safety and reliability~\cite{sotif2022}. Similarly, the recent EU regulation for type-approval of autonomous vehicles states that autonomous vehicles should be able to perform the driving task in all reasonably foreseeable, critical scenarios in the operational environment, which include not only those observed from the data, but also the possible ones derived or extrapolated from the scenario description~\cite{eu2022}. 

Identification of critical scenarios for testing autonomous driving systems is an emerging industrial need~\cite{lou2022testing}. 
Even though critical scenario identification has received increasing interest in academic research, most studies, as surveyed and presented by Zhang et al., focus on developing and demonstrating principles of critical scenario identification approaches~\cite{zhang2022finding}, 
and very few industrial applications of such approaches have been presented so far~\cite{reisgys2022scenario}. Current industrial research mainly present the tool chain, e.g., Siemens et al.~\cite{simens2022whitepaper} and Hallerbach et al.~\cite{hallerbach2018simulation}, rather than implementation, practices, and performance within the organization. As a result, the industry practices and challenges for critical scenario identification  are not well explored and understood. 
Consequently, it is hard for academia to know whether their approaches are feasible and effective when implemented on an industrial scale, whether the identified critical scenarios are realistic, and whether they contribute to the testing and safety analysis to the extent expected. 

In this study, we explore industry practices and the challenges for critical scenario identification for testing autonomous driving systems. Specifically, we address two research questions: 
\begin{enumerate}
    \item [RQ1]
\textit{-- What are the industry practices related to critical scenario identification and testing}, and 
\item[RQ2] \textit{-- What are the challenges related to critical scenario identification and testing}. 
\end{enumerate}We conducted 10 interviews with 13 interviewees from 7 companies in Sweden, where the companies are intensively involved in and contributing to the advancement of autonomous driving. The interviewees are working or have worked with the testing of autonomous driving systems. During the interviews, we inevitably discussed substantial practices and challenges for testing of autonomous driving systems and scenario-based testing in general. In this study, we particularly focus on analyzing and presenting the findings in relation to \emph{critical} scenarios, including their definitions, selection criteria, identification approaches, related tools and platforms, validation and evaluation, and some general initiatives, while leaving out general aspects of scenario-based testing. In addition, challenges such as insufficient definition, approaches hard to implement, assessment criteria hard to define, and insufficient tools and platforms, were also identified. We conclude that there is no single universal approach for critical scenario identification, and different approaches must be combined. Also, different stakeholders must collaborate and continuously learn and improve the testing techniques for autonomous driving.    

The rest of the article is organized as follows: in Section~\ref{sec:related_work_terms}, we describe the relevant literature and terms used in this study. We formulate the methodology in Section~\ref{sec:method}, and present the results of the study in Section~\ref{sec:results}. In Section~\ref{sec:discussions}, we discuss our analysis and insights based on the results, and we also discuss the validity of the study. Lastly, we conclude the article in Section~\ref{sec:conclusions}. 

\section{Background and Related work}
\label{sec:related_work_terms}

In Section~\ref{sec:related_work}, we present some studies related to critical scenario identification, and in Section~\ref{sec:terms}, we describe the terms and concepts that we use in this study. Even though the literature presented does not study industry practices on critical scenario identification and testing, they provide a very good basis and input for the interviews with respect to different critical scenario identification approaches, testing techniques, and testing practices for autonomous driving systems. 

\subsection{Related Work}
\label{sec:related_work}


Ponn et al. describe critical scenario identification as selecting, executing, and evaluating a scenario with relevant criticality metrics~\cite{ponn2020identification}. According to Zhang et al., it is vital to understand what is critical and how to select the critical scenarios since the potential number of possible scenarios in the operational environment is infinite and testing all of them is unfeasible~\cite{zhang2022finding}. Given that, different approaches are proposed for identifying critical scenarios, e.g., using experts, naturalistic driving data, or guided search, and the identified critical scenarios are used for testing autonomous driving systems to validate their safety and reliability~\cite{song2022critical}.

Zhang et al. performed a systematic literature review on critical scenario identification and provided a taxonomy for the approaches based on 86 papers from 2017--2020~\cite{zhang2022finding}. The taxonomy contains four categories of approaches, which we summarize as follows: (1) critical scenario exploration without parameter trajectory, using, for example, search-based algorithms. The general principle of these approaches is to identify parameters of interest from the operational design domain and explore the critical scenarios using guided search and simulation. An example is Song et al., where critical scenarios are optimized based on surrogate measurements like TTC (Time-to-Collision)~\cite{song2022critical}, (2) critical scenario exploration with parameter trajectories, where parameters can change during the scenario execution instead of following a constant pattern. An example is Karunakaran et al., who used reinforcement learning to find the critical scenarios in a pedestrian crossing situation~\cite{karunakaran2022critical}, (3) induced learning that derives critical scenarios from existing data sources. For example, Gambi et al. reconstructed accident reports using natural language processing and simulated them for testing autonomous driving systems~\cite{gambi2019generating}, and (4) deductive learning, which uses domain expertise and ontology to derive critical scenarios. For example, Ponn et al. recruited experts from the automotive industry to support critical scenario creation~\cite{ponn2020identification}.
        
Riedmaier et al. surveyed articles on scenario-based testing in general, rather than critical scenarios only~\cite{riedmaier2020survey} and presented a taxonomy of scenario-based approaches for the safety assessment of autonomous vehicles. The taxonomy includes: (1) knowledge-based approaches, which leverage expert knowledge or existing ontology to derive scenarios, such as Li et al.~\cite{li2020ontology}, (2) data-driven approaches where scenarios are identified, extracted, created, or clustered, based on data in different forms, such as sensory data collected from naturalistic driving, accident reports, or synthetic data from the simulation, like surveyed by Cai et al.~\cite{cai2022survey}, and (3) search-based approaches that generate scenarios by searching in the scenario space, using, for example, search-based algorithms, like Song et al.~\cite{song2022critical, song2021industrial}. While the scope of this article is broader than the one by Zhang et al.~\cite{zhang2022finding}, the taxonomies are still consistent and can clearly relate to each other. In this study, we adopt the taxonomy from Riedmaier et al.~\cite{riedmaier2020survey} when describing the field to our interviewees, as it is more general and intuitive, but still focuses on critical scenarios. 

Lou et al. performed a survey to explore the industry practices of testing autonomous driving systems and to what extent the existing test techniques studied in academic research address the industrial needs~\cite{lou2022testing}. In their study, they interviewed 10 industry practitioners in the autonomous driving domain, and collected 100 online surveys to consolidate more insights from this field. As a result, they identified seven common practices for testing autonomous driving systems, regarding such as sources of test scenarios, simulation testing, on-road testing, etc. They also identified four emerging needs, e.g., identifying critical scenarios (which they framed as ``possible corner cases and unexpected driving scenarios''), and tool support for labelling data and creating complex driving scenarios. Subsequently, they reviewed the literature on this topic and developed a taxonomy of the research, and analyzed the gaps between the emerging needs and current testing techniques explored in the literature. 
In this study, we focus on industry practices and challenges for identifying \emph{critical} scenarios for testing autonomous driving systems, specifically.  

Industrial studies on critical scenario identification are also reported. For example, Siemens presented a white paper describing the tool chain and concept of critical scenario identification using a search-based approach~\cite{simens2022whitepaper}. Hallerbach et al. is another example of such studies that demonstrates the principle and the tool sets for identifying critical scenarios from Opel~\cite{hallerbach2018simulation}. While both are industry reports, they focus more on demonstrating their tools and integrating them for identifying critical scenarios. In contrast, we study the practices of how critical scenario identification is implemented and how critical scenarios are used for testing in the industry. 

\subsection{Terms and concepts}
\label{sec:terms}

In this section, we clarify some general terms and concepts that are used in this study, which may entail different meanings or interpretations. Among them, \textit{scenario} is the preliminary term we use throughout the study adhering to the definition by Ulbrich et al.~\cite{ulbrich2015defining}. They define a scenario as a temporal sequence of scenes, and each scene consists of all stationary objects like road infrastructures, constructions, occlusions, etc., and dynamic objects like different road users (e.g., pedestrians, cyclists, vehicles), as well as their actions (e.g., left turn, cut-in, or cut-out) and events (e.g., congestion by other vehicles), associated with this scene. When it comes to critical scenario identification, we adhere to the definition of \emph{concrete scenarios} from Menzel et al., in which a scenario is defined as concrete values for all relevant parameters selected for this scenario, e.g., speed, and acceleration of the vehicle, number of lanes, weather, etc.~\cite{menzel2018scenarios} 

The \textit{critical scenario} is another primary term as it is the focus of our study. We observed in our previous study~\cite{song2022critical} that there are different definitions and interpretations of critical scenarios (e.g., hazardous scenarios, challenging scenarios, complex scenarios, or corner cases). However, they are similar, and most of them are concerned with unsafe situations or consequences like collision or near-collision. Zhang et al. describe critical scenarios as ``scenarios that cause potential risks of harm, which need explicit consideration for risk investigation and potential mitigation measure''~\cite{zhang2022finding}. The recent EU regulation defines critical scenarios as unexpected conditions (e.g., traffic, weather) that require emergent actions~\cite{eu2022}. 
Besides, we adopt and use the term \textit{nominal scenarios} from the latest EU regulation to represent the regular (reasonably foreseeable) scenarios that involve non-critical interactions and require no immediate and emergent actions for autonomous driving systems~\cite{eu2022}.

We use \textit{critical scenario identification} (CSI) to refer to the process of finding the critical scenarios, using different approaches. In previous studies, it is often described as searching, creating, generating, optimizing, or extracting critical scenarios. To avoid using different terms and ambiguities, we stick to the term \textit{identification} unless there is any particular perspective of the approach that we want to stress (e.g., optimization is occasionally used in search-based approaches). It reflects the purpose of identifying critical scenarios and includes different forms as just mentioned. The \textit{operational design domain}, as defined by Gyllenhammar et al.~\cite{GyllenhammarODD20}, and different standards, like SOTIF~\cite{sotif2022}, J3016 from SAE International~\cite{j30162021}, and EU-2022/1426~\cite{eu2022}, refer to the operational environment where the function is supposed to work, characterized by, for example, weather, time of the day, country or region, road types (e.g., highway, or urban streets), speed limits, etc.

Lastly, we employed an open and inclusive definition of \textit{autonomous driving systems}, which can be used for ADAS (Advanced driver assistance systems), autonomous driving systems, or active safety systems, as long as they perform or support the autonomous driving tasks to some extent. The reason is that our interviewees may work on different autonomous driving systems at different levels of automation (see levels 0--5 from SAE International), and have different views of autonomous systems. The same system (e.g., for parking) can be implemented and defined at different levels of automation by SAE International (see examples and full description in J3016~\cite{j30162021}). Unless the interviewees explicitly mention a specific level or use a specific term like ADAS, it is hard to tell at which level of automation the system is. Besides, if we limit the specific definition and examples of autonomous driving systems, the interviewees may not find specific projects or systems complying with our definition, thus limiting the findings of the study. The point is that we want to be inclusive when exploring industry practices, and we do not want to restrict our study to autonomous driving systems at a specific level or with any particular feature.

We recognize there could be more relevant terms being used, such as fault, failure, hazard, malfunctioning behavior, functional safety, etc. We do not explain every term here, but refer to industrial standards like ISO-26262~\cite{iso262622018}, SOTIF (ISO/PAS-21448)~\cite{sotif2022}, J3016 from SAE International~\cite{j30162021}, or studies like Zhang et al.~\cite{zhang2022finding}, for a full list of terms and concepts related to autonomous driving. 

\section{Methodology}
\label{sec:method}

The study is an interview study, in which we mainly used semi-structured interviews~\cite{runeson2009guidelines} to gather insights from industry practitioners.  
The implementation of the interviews was guided by the suggestions from Rowley~\cite{rowley2012conducting} and Runeson et al.~\cite{runeson2009guidelines} as well as our experience from previous studies~\cite{song2021concepts}. After the interviews, we employed thematic analysis~\cite{cruzes2011recommended} to synthesize the data and to model the concepts in a structured and coherent way. The study consists of three stages, \emph{pre-interview, in-interview}, and \emph{post-interview}, as detailed in the following sub-sections. 

\subsection{Pre-interview}

\subsubsection{General preparation}
\label{sec:preparation}

We analysed relevant literature, including both primary studies and secondary studies, as discussed in Section~\ref{sec:related_work}. We also studied some commercial reports (supplier name shall not be revealed) and public test data for autonomous vehicles (e.g., public road testing mileage, disengagement, and collision report from DMV California~\cite{dmvcalifornia}
) to expand the potential sampling pool for the interview. After that, we presented our study design in two academic workshops to receive feedback and suggestions. 
Those workshops provided us with potential interview candidates, concerns on threats to validity, interview guidelines, comments for the design of the study, and ethical considerations. 
In addition, we discussed threats to validity in general at a software engineering research group (SERG~\footnote{https://serg.cs.lth.se/}) seminar 
and evaluated our study using different frameworks as described in Wohlin et al.~\cite{wohlin2012experimentation}, Maxwell~\cite{maxwell1992understanding}, and Storey et al.~\cite{storey2020software}. The discussion on the validity of the study is detailed in Section~\ref{sec:threats}.

\subsubsection{Participant selection} 

Participants for the interviews were selected mainly in two phases. 
Initially, we used a convenient sampling approach where participants were selected mostly for a previous collaboration with us, easy accessibility, and geographical closeness~\cite{baltes2022sampling, ghazi2018survey, etikan2016comparison}. 
To expand and consolidate the insights from the interviewees, 
we used a snowballing approach 
to get more candidates from our interviewees. We also used purposive sampling to approach participants based on their focus area and competence~\cite{baltes2022sampling, etikan2016comparison}, until we sampled most companies and roles that we know in Sweden and we expect similar responses from further interviewees. 


\begin{table}[htbp]
\centering
\small
\begin{tabular}{|l|p{0.25\textwidth}|l|p{0.35\textwidth}|l|}
    \hline 
    \# & Role & Company & Responsibilities & Experiences \\
    \hline
    P1  & Post-doc \& Functional Safety Engineer &  C1 & Research projects, functional safety, distributed ML systems & 5--10 years \\
    \hline
    P2 & Business Developer & C2 & Business development & 10--15 years \\
    \hline
    P3 & Head of Architecture & C3 & Software architecture for ML, including AD & 25--30 years \\
    \hline
    P4 & Product Owner & C4 & Perception for low-speed maneuvering domain of AD & 15--20 years \\
    \hline
    P5 & AD Specialist & C5 & AV testing and safety strategies & 10--15 years \\
    \hline
    P6 & Research Engineer & C6 & AD research projects & 1--5 years \\
    \cline{1-2}   \cline{4-5}
    P7 & Test Leader &  & Software verification and validation & 20--25 years \\
    \hline
    P8 & Solution Architect & C4 & Sensor performance & 20--25 years \\
    \hline
    P9 & Validation Engineer & & Verification and validation of AD & 5--10 years \\
 \cline{1-2}   \cline{4-5}
    P10 & Development Engineer & C7 & ML tool-chain, verification and validation of AD & 10-15 years \\
    \hline
    P11 & Senior Engineer & C7 & Virtual testing platform & 10--15 years \\
    \hline
    P12 & Product Owner & & Verification, validation, and safety argumentation of AD &  5--10 years \\
 \cline{1-2}   \cline{4-5}
    P13 & Senior Engineer & C4 & Verification, validation, and safety argumentation of AD & 10--15 years \\
    \hline
\end{tabular}
\caption{List of interviewees selected and participated in the interview. Merged cells in the Company column indicated interviews conducted with pairs of interviewees. The Experiences column indicates the total number of years' professional employment of the interviewee.}
\label{table:interviewees}
\end{table}

\begin{table}[htbp]
\centering
\begin{tabular}{|l|l|l|}
    \hline 
    Company & Domain & Size \\
    \hline
    C1 & Motor Vehicle Manufacturing & 10~001+ \\
    \hline
    C2 & IT Services and IT Consulting & 2--10 \\
    \hline
    C3 & IT Services and IT Consulting & 1001--5000 \\
    \hline
    C4 & Motor Vehicle Manufacturing & 10~001+ \\
    \hline
    C5 & Motor Vehicle Manufacturing & 1001--5000\\
    \hline
    C6 & IT Services and IT Consulting & 201--500 \\
    \hline
    C7 & Motor Vehicle Manufacturing & 10~001+ \\
    \hline
\end{tabular}
\caption{List of companies that participated in the interview. The Size column indicates the total number of employees of the company, which is taken from their LinkedIn profiles.}
\label{table:companies}
\end{table}

In total, we selected 13 interviewees (see Table~\ref{table:interviewees}) from 7 different companies in or related to the autonomous driving domain in Sweden, as shown in Table~\ref{table:companies}. 
We adopted the company profile (i.e., domain and size) from LinkedIn. Similarly, we retrieved the experiences info of the participants from their LinkedIn profiles, where only full-time professional employment was counted. Overall, our interviewees cover various roles and responsibilities, and have substantial experiences (9 out of 13 had over 10 years of experience), although not all of them were dedicated to autonomous driving. Also, all of them were involved or had a background in the testing of autonomous driving systems. To avoid mentioning or referring to specific projects, teams, or products, we provide a summary instead of expanding all responsibilities and tasks for each interviewee. 

\subsubsection{Participant consent form}

A participant consent form (see Zenodo~\cite{qunying_song_2023_7798703}) was created 
to include statements of voluntary participation in the study, the purpose of the study, and information that they could withdraw at any time or refuse to answer any questions; that the interview could be recorded, transcribed, and analysed for research purpose, and will be anonymized and treated confidentially. The participant consent form was distributed before the interview, and all participants agreed to it. Among them, 9 of 13 provided a signed copy of the participant consent form. Still, we repeated the form and got consent from the participant before starting the interview.

\subsubsection{Interview questions}
The interview script has nine questions (see Zenodo~\cite{qunying_song_2023_7798703}) that were designed and divided into three parts, as summarized in Table~\ref{table:interview_questions}. We reviewed the questions to ensure they were not hard to understand or too intrusive. To avoid conflicting with company confidentially, we aimed at getting general views from the interviewees rather than asking anything concrete about their work or organizations. 

\begin{table}[htbp]
\centering
\begin{tabular}{|p{0.95\textwidth}|}
    \hline Part 1 -- Who are you? \\
        1) What is your role in your organization? \\ 
        2) What is your main focus, especially in relation to autonomous driving? \\
        3) How have you been involved in testing autonomous driving systems? \\
    \hline Part 2 -- What do you think of using critical scenario identification for testing? \\
        4) How do the critical scenario-based and other testing approaches complement each other? \\
        5) What are the general advantages and limitations of critical scenario-based approaches? \\
        6) How to improve the relevance of generated critical scenarios for testing? \\
        7) Which approach is most feasible from a practical view? \\
    \hline Part 3 -- What else to complement? \\
        8) What are the open challenges for testing autonomous driving systems, particularly in relation to the use of critical scenario identification? \\
        9) What are your general feedback and further interviewee candidates to recommend?\\
    \hline
\end{tabular}
\caption{A brief version of the interview questions, see Zenodo~\cite{qunying_song_2023_7798703} for the full version}
\label{table:interview_questions}
\end{table}

\subsection{In-interview}

\subsubsection{Pilot interview}

As suggested by Jennifer et al., a preliminary pilot interview (with P1) was conducted to make sure that the interviewees could understand our questions~\cite{rowley2012conducting}. Based on the pilot interview, we reflected on whether the questions were clear and explicit. In some cases, the interviewee wanted us to repeat or further explain the question, which was a clear indication that the question was not clear in the first place. Also, we tried to make sure the questions were not too general. For example, the interviewees may have various interpretations of the question and propose responses that were irrelevant or did not explicitly address the core of the question. Lastly, we also evaluated whether any question could be too intrusive and the interviewee may refuse to answer due to confidentiality concerns over the company's secrecy.

All of us (the three authors of this study) attended the pilot interview online via Microsoft Teams. 
The interview was considered successful and smooth, as reviewed by the study team, where the interviewee could understand our questions and gave us insightful responses as well as expanding them for further discussions. 
Overall, we followed the protocol as described in Section~\ref{sec:protocol}, and kept the questions to be more general, open, and friendly to the interviewees, yet still maintaining relevance to the topics of the study and can be extended based on how the interviewee reacts.

\subsubsection{Interview protocol}
\label{sec:protocol}
In total, we conducted 10 interviews, including the pilot interview. As shown in Table~\ref{table:interviews}, all interviews were conducted online and lasted from 55 to 75 minutes. Three interviews (No. 6, 8, and 10) had two interviewees from the same company participated together. Except for the pilot interview, which all three authors attended, the other interviews were conducted by two authors each. The first author (A1) participated in all interviews and acted as the main driver, and either the second (A2) or third author (A3) assisted with the interview. Microsoft Teams was used since some companies had restrictions using other online tools like Zoom, and an auto-transcription function is provided in Teams, which saves a lot of time compared to manual transcription. Overall, the interview contains mainly four parts -- a short introduction of the study, a brief overview of the research studies in the field, a walk-through of the prepared questions, and conclusion remarks. 

\begin{table}[htbp]
\centering
\begin{tabular}{|l|l|p{0.18\textwidth}|p{0.1\textwidth}|l|l|}
    \hline 
    Interview & Participants & Length (rounded) & Form & Interviewers & Date \\
    \hline
    1 (Pilot) & P1 & 60 min & Online & A1*, A2, A3 & 2022-05-12 \\
    \hline
    2 & P2 & 65 min & Online & A1*, A3 & 2022-06-15 \\
    \hline
    3 & P3 & 55 min & Online & A1*, A3 & 2022-08-23 \\
    \hline
    4 & P4 & 65 min & Online & A1*, A2 & 2022-08-26 \\
    \hline
    5 & P5 & 65 min & Online & A1*, A2 & 2022-08-30 \\
    \hline
    6 & P6, P7 & 60 min & Online & A1*, A3 & 2022-09-16 \\
    \hline
    7 & P8 & 65 min & Online & A1*, A2 & 2022-09-19 \\
    \hline
    8 & P9, P10 & 75 min & Online & A1*, A3 & 2022-10-31 \\
    \hline
    9 & P11 & 65 min & Online & A1*, A3 & 2022-10-31 \\
    \hline
    10 & P12, P13 & 55 min & Online & A1*, A2 & 2022-11-09 \\
    \hline
\end{tabular}
\caption{Interview list - Interviewers refer to the three authors in the byline who participated in the interview (e.g., A1 means the first author), and the asterisk (*) sign indicates the main driver of the interview }
\label{table:interviews}
\end{table}

\begin{enumerate}
    \item Introduction -- a quick introduction of ourselves, purpose of the study, participant consent form, expected duration and agenda of the interview, and ask permission to record a video for the interview. This part is planned for 5 minutes.
    \item Research overview -- start by asking what is their interpretation of critical scenarios in the context of testing autonomous driving systems, then present the commonly-adopted definitions of scenarios, i.e., Ulbrich et al.~\cite{ulbrich2015defining}, Menzel et al.~\cite{menzel2018scenarios}, related standard SOTIF~\cite{sotif2022, simens2022whitepaper}, and taxonomy of approaches for critical scenario identification and example studies of them, including, knowledge-based, data-driven, and search-based approaches~\cite{riedmaier2020survey}. Lastly, ask whether the interviewees agree with the concepts and terms or any questions or comments they have. This part is planned for 10-15 minutes.
    \item Question list -- walk-through the questions in a semi-structured manner~\cite{runeson2009guidelines}, which means adaptations and explanations can be made, and new questions can be added. We also give a brief recap to ensure we understand the answers the same way as the participants. Also, we put our questions and lead the discussion in a way to set some kind of basis/context for the next question, if possible, so that the entire interview is more progressive. Initially planned for 30-40 minutes, depending on the engagement from the participants. 
    \item Conclusion remarks -- acknowledge the participation of the interviewee(s), inform the coming steps, say we will review the transcription and fix errors if occurred, highlight the unclear or uncertain parts that need further clarification and send the transcription back to them. The planned time is less than 5 minutes for this part.
\end{enumerate}

\subsection{Post-interview}
\subsubsection{Transcription proofreading} We used a semi-automatic transcription approach where auto-transcription from the software (Microsoft Teams) was taken as a basis, and manual correction was made to improve the quality of it. Before we formally started each interview, we asked and obtained permission from the interviewee(s) to record the interview and enabled the auto-transcription option. After the interview, a video file (in .mp4 format) and a transcription file (either in .docx or .vtt format) were automatically generated by Microsoft Teams and saved by us locally. 

The initial transcription file was already formatted into text snippets chronologically by Microsoft Teams. Each snippet is associated with: (1) a starting timestamp and an ending timestamp, (2) name of the speaker, and (3) the text transcribed from the speaker during this time period. In the first iteration, the first author listened to the recorded video of each interview and confirmed the text in the transcription. Overall, the auto-transcription was accurate. Nevertheless, errors still occurred, for example, one or few words being transcribed as something else that is similar, and thus were revised. It happened, though rarely, that auto-transcription was missing for a short duration due to possible technical issues in the software or network, so it has to be compensated manually. If any places were identified as unclear and unable to correct based on the video, they were highlighted in the transcription and marked "further clarification needed".

Also, there were many filler words (e.g., uhms and aahs) and repetitions from the speakers, which are not crucial for interpretation and analysis, and thus were removed. In some cases, the speakers started with some unrelated phrases or terms in a grammatically incorrect way because they were not fully prepared yet to answer the question but later switched their minds and rephrased their speech. In those situations, the first part was removed, which was unnecessary and could mislead the interpretation. Overall, we used a Clean Transcription (also known as edited transcription
) approach where filler words, stutters, and repetitions should be removed. In comparison, we considered Verbatim (also called literal transcription, transcribing everything, even pauses, facial expression, and body language etc.~\cite{mclellan2003beyond, halcomb2006verbatim}) and Summary (summarizing while editing~\cite{mclellan2003beyond}) approaches are either too fine-grained or coarse-grained, especially the summary transcription approach which could be inaccurate and too condensed.

In the second iteration, the first author reviewed the revised transcription with the playback of the recording again, but to see whether there were any parts or places that were hard to understand or not entirely sure about what the interviewee meant, these places were also highlighted and marked as "further clarification needed". Note that in this iteration, we did not aim to analyse, but to ensure the transcription is comprehensible to us. Next, the transcription was sent back to the interviewees, and they could review and edit their answers if they wanted. Confirmation was requested if we needed support from them to clarify or understand certain parts (marked with further clarification needed). Finally, the transcription was completed if we were fully aligned with the participants and no unclear parts were left. 

\subsubsection{Thematic analysis}
\label{sec:thematic_analysis}
We used a shared workspace in Microsoft Teams to collaborate on the coding and thematic modeling of the interview data. We uploaded the transcription files of the interviews to the shared workspace where we could open and edit the transcription (in .docx format) files directly. In the application, we could view, comment, and edit the same file simultaneously for coding purposes. 
We followed the guidelines from Cruzes and Dybå~\cite{cruzes2011recommended} and conducted the thematic analysis and modeling in four steps as follows:

\begin{enumerate}
    \item Primary coding exercise (with 3 transcripts) -- All authors independently coded one transcript per author to generate explicit and standalone codes, which one can understand without re-visiting the full text. Important text segments addressing or related to our research questions were selected (extract data) and commented (code data), where the comment was the proposed code. Next, we reviewed each others' transcripts and proposed changes or added thoughts by replying to the codes. Then, we organized a group exercise to review and discuss the codes for each transcript until reaching a full agreement. 
    \item Primary modeling exercise -- All authors independently translated codes from step (1) into themes, and proposed hierarchical models of themes and codes. 
    The top-level themes are the research questions. We then 
     iteratively merged our models into one model in several rounds of group exercises until we thought the model was coherent and mature. 
    \item Continued coding (with 7 transcripts left) -- The first author continued coding the remaining transcripts by following the same principles as agreed in the primary coding exercise. 
   When the first author was not confident with the interpretation or did not find suitable code for a specific transcript or part of the transcript, the other authors were consulted. 
    \item Continued modeling -- The continued modeling was conducted in two steps. First, the first author added and merged the codes from step (3) into the primary model. Second, we (all three authors) organized two model review sessions (1.5 hours each) to walk through the codes and themes one by one. Each code and theme was reviewed, elaborated, and discussed both from a microscopic view (whether the code was grouped in a reasonable theme) and a macroscopic view (whether the themes on the same level and different levels form a coherent and consistent model). As a result, we developed the final model (the final model is available on request for now, and will be published publicly on Zenodo if the paper gets accepted).
\end{enumerate}

\section{Results and Analysis}
\label{sec:results}

In this section, we present the results from the interviews and analysis based on that. Specifically, we demonstrate \textit{RQ1 -- Practices of testing with CSI} in Section~\ref{sec:practices} and \textit{RQ2 -- Challenges for testing with CSI} in Section~\ref{sec:challenges}. We separate the \emph{practices} (i.e., Section \ref{sec:definition} -- \ref{sec:tools}) from \emph{initiatives} (i.e., Section \ref{sec:initiatives}) to avoid the ambiguities of interpreting things that the interviewees proposed or suggested, yet not implemented as the current practices of the industry. While initiatives are also part of the practices we identified, they are more theoretical proposals to guide and inspire future practices, but the suggestions and proposals may not be proven yet. When discussing practices, we primarily focus on what has happened, e.g., observations, real examples, and things that are used or implemented. Also, we use the form, e.g., P6-7, referring to the two participants (P6 and P7) from the same organization who joined the same interview and contributed together in one conversation.

\subsection{Practices of testing with CSI (RQ1)}
\label{sec:practices}
In this section, we present the practices of testing used or implemented in the industry, with respect to critical scenario identification. During the interviews, we inevitably touched upon practices of testing autonomous driving systems and scenario-based testing in general. In this study, we mainly focus on practices related to critical scenarios and critical scenario identification approaches. 

\begin{figure}[htp!]
\includegraphics[width=\textwidth]{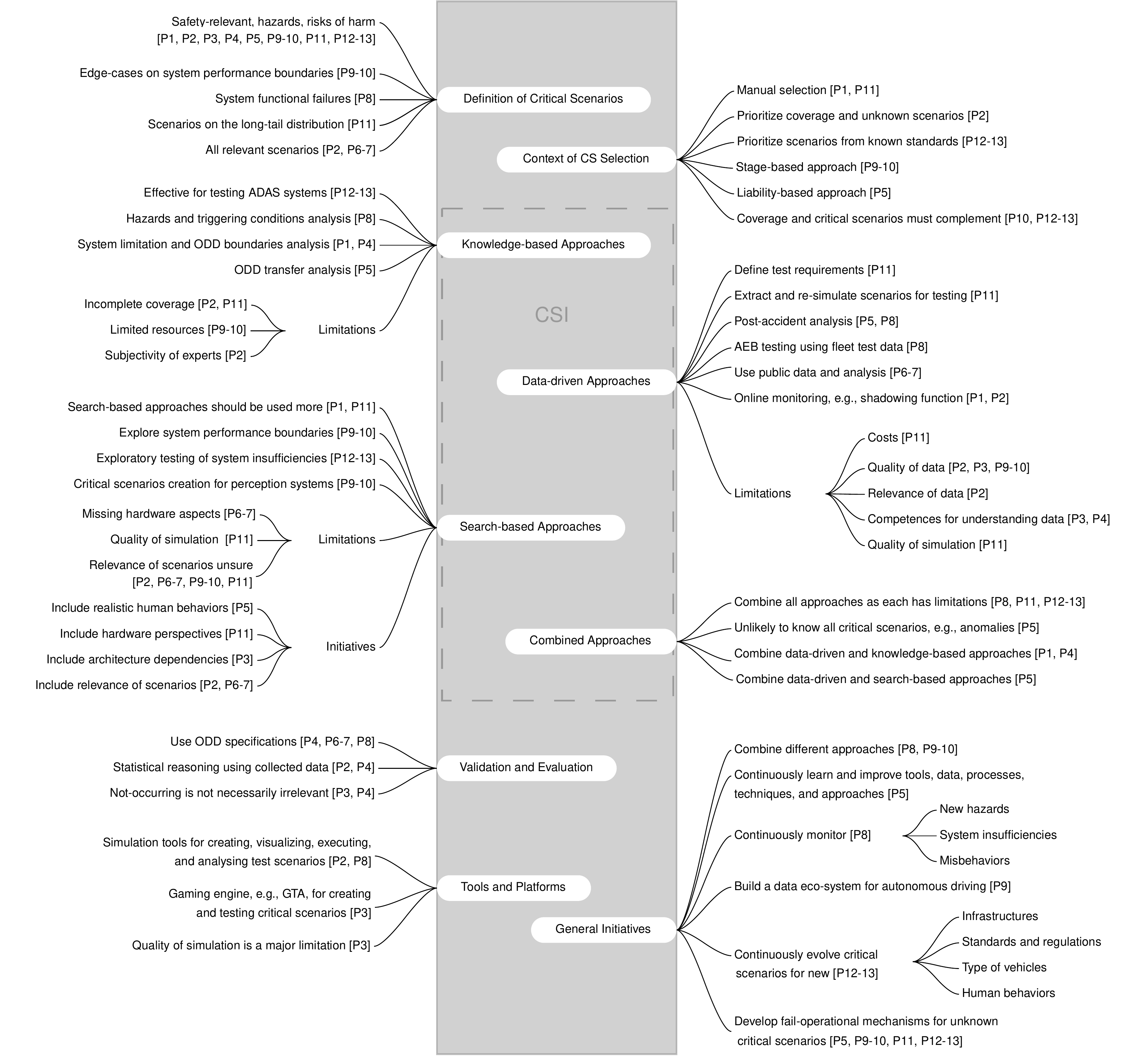}
\caption{A summary thematic model of (RQ1) practices for critical scenario identification and testing. The model includes nine top-level themes in the grey rectangle for critical scenario identification and testing, and sub-level themes which are mostly implementations and examples, observed limitations, and proposed initiatives. Specifically, the four themes in the dashed rectangle are four types of approaches for CSI (critical scenario identification). The CS is the acronym for critical scenarios.}
\label{fig: practices}
\end{figure}

We present a summary thematic model for the practices in Fig.~\ref{fig: practices}, and elaborate the (nine) top-level themes in the grey rectangle further in Sections~\ref{sec:definition}--\ref{sec:initiatives} respectively, in top-down order as in Fig.~\ref{fig: practices}. Specifically, the four themes in the dashed rectangle are four types of approaches for CSI (critical scenario identification). The model is summarized from the full thematic model of the study, and includes the nine top-level themes under RQ1 and the sub-themes on the appropriate level of abstraction. The sub-themes are embodied with minimum explanation and context from the codes and interview transcripts to be self-explanatory. 

\subsubsection{Definition of critical scenarios}
\label{sec:definition}

The most common definition of critical scenarios (used by 10 out of 13 interviewees) refers to critical scenarios as \textit{safety-relevant} and scenarios that can expose harm or cause potential hazards for autonomous vehicles and other road users. While the concept remains the same, P9-10 indicated that they use \textit{edge cases} instead of critical scenarios within their organization, i.e., they are interested in identifying the system performance boundaries, not only those unsafe scenarios. 
Nevertheless, P9 still provides two interpretations of critical scenarios in relation to safety, namely qualitative and quantitative. The qualitative definition implies a qualitative analysis and assessment of certain hazards, including their characteristics, triggering conditions, and the causal relationships between the hazardous events and the triggering conditions. In contrast, the quantitative definition implies using metrics to evaluate the potential criticality of scenarios, such as risk estimation, probability, or severity. 

There are other interpretations of critical scenarios from the interviewees. Interviewee P8 puts critical scenarios in the context of sensor performance, which is the main responsibility of their team and claimed any scenarios that result in \textit{functional failures} are deemed as critical scenarios. For example, a failure of a sensor detection of an object or under a specific circumstance would be critical to investigate and test. P11 from another company (C7) provides yet another dimension to understanding critical scenarios, and mentioned scenarios on the \textit{long-tail distribution} of all scenarios are critical. In other words, critical scenarios are those that are uncommon and have an extremely low probability of occurrence. 
Interestingly, P6-7 expressed that \textit{all relevant scenarios} are critical, especially when the technologies for autonomous driving and testing such systems are immature. Similarly, P2 articulated that a scenario can be critical in terms of \textit{relevance and necessity} for testing, like a simple but common scenario, or it is critical with respect to the \textit{system performance}, for example, collision or near-collision. ``The problem is we don't know if a scenario is critical or not until we test it, and in many cases, we don't fully understand how the changes are related to the test scenarios, so we have to test them all'', and for that reason, P2 insisted that all relevant scenarios should be considered critical for testing.

\subsubsection{Context of critical scenario selection} 

In general, the selection of test scenarios, including both nominal and critical ones, is still dominantly \textit{manual}. In C1, according to P1, specific test teams consisting of test engineers are responsible for analysing the systems, designing the testing strategies, and selecting test scenarios to perform. As a recent initiative in C7, they attempt to include more roles other than testers to create and select test scenarios, like developers, as they know better the system design and implementation. Since critical scenarios constitute an important subset of the test scenarios, it is imperative to understand when and how they are used for testing.

P2 prioritized exploring more \textit{unknown scenarios} in the test and emphasized that we should not overtly favor the critical scenarios. As they said: ``critical scenarios can be those that rarely happen in real-world traffic, and we need to test more relevant scenarios in order to demonstrate the safety of autonomous vehicles. Thus, improving the coverage of test scenarios is essential and will naturally include critical scenarios.'' P12-13 shared a similar opinion that they prioritize more on the scenarios that fall into the \textit{known standards}, e.g., SOTIF, and ensure those scenarios are covered in the test. However, they added that it does not mean scenarios not defined in the standards should not be tested, but receive less effort. P8 gave an example of where critical scenarios were selected for testing sensor detection range in C4. The focus was to identify and test the worst-case scenario and assume scenarios below the worst case would be safe.

Critical scenarios and scenario coverage are debated in relation to the completeness of testing and long-tail issue, as described by P9-10. As a pragmatic solution, they used a \textit{stage-based approach} to determine when critical scenarios are selected for testing. In the early stage of development, it is unnecessary to focus too much on critical scenarios, as even the most simple and common scenarios might fail. In the middle stage, the testing mainly aims to validate specific properties or features of the most recently developed functions, e.g., perception; thus, critical scenarios are still not the major focus. In the mature stage, the system gets more stable and can handle most common scenarios, and testing resources can be placed on the critical scenarios. As one may question the division of different stages, they have confirmed it is experience-based. In general, the early stage intends to build a backbone or prototype without having the full requirements, finalized architecture and design, and relevant data. The middle stage entails major functional modules being developed and integrated, and the system can handle some known scenarios. The mature stage includes testing the full vehicle on public roads like Waymo does, and it is important to test some complex and extreme scenarios, e.g., abrupt cut-in, inclement weather, or pedestrian joy-walking. 

In contrast to other interviewees, P5 presented their thoughts from a \textit{liability} perspective and claimed that the focus of testing and selection of test scenarios are also subject to the liabilities of the organization. Autonomous driving function suppliers (e.g., those who develop autonomous driving algorithms) have more limited responsibilities, so they might focus on testing and mitigating problematic scenarios rather than finding all possible scenarios. However, autonomous vehicle suppliers (e.g., those that manufacture the entire vehicles) have full responsibility for their vehicles, especially safety. They have to ensure that the vehicles can handle all scenarios in their operational environment and remain safe, so only testing critical scenarios is not enough, but all relevant scenarios.

Eventually, covering more possible scenarios and identifying critical scenarios must be complementary in the test. As said by P12-13: ``we need to identify more unknown scenarios, especially those hazardous ones. Still, we also need to define and evolve a scenario catalog or database where each software release must be tested. The reason is that if we just rely on the critical scenarios, we would miss a lot of problems that can occur in the real world, so we need to cover the known and common scenarios as well.'' P10 gave an example of offline testing for machine learning systems, i.e., perception, where they started by aiming for higher coverage of scenarios (images), then the critical scenarios (edge images), which are very challenging for the system to classify correctly.

\subsubsection{CSI approaches -- Knowledge-based}
\label{sec:knowledge-based}

Knowledge-based approaches refer to deriving critical scenarios using expert knowledge or any previously established ontology in the domain, e.g., laws, standards, regulations, etc. As articulated by P6-7: ``prior knowledge and expertise learned from different sources, e.g., certain data statistics, is always used for creating test scenarios, even without referring to them explicitly.'' P12-13 mentioned that knowledge-based approaches are especially useful for testing ADAS systems. However, P8 expressed concerns about using such approaches for testing high-level autonomous driving systems, where a sheer number of hazardous events may occur, so they strive to minimize the unknown scenarios in testing in a way like SOTIF proposes.

Several examples of using knowledge-based approaches were given in the interviews, 
P8 described hazard analysis as part of their testing activities, where experts try to list all known insufficiencies of the functions, acceptance criteria, and triggering conditions (critical scenarios) for the hazards. P1 described identifying critical scenarios based on system limitations and boundaries of the operational design domain (ODD). Generally, they (P1) first clearly define a system's ODD, and nail down what is expected in the ODD and what is not; then, they draw the boundaries and create critical scenarios based on that. P4 used the same principle for identifying critical scenarios for, e.g., mono depth estimation issues with one camera. In addition, P5 discussed that switching from one ODD to another is a grey area and can certainly expose more critical scenarios.

Even though knowledge-based approaches are beneficial, and are said to dominate the current scenario creation for P11, limitations were observed and discussed by the interviewees. The major limitation is the coverage of critical scenarios (indicated by P11) and, consequently, the completeness of critical scenarios tested (raised by P2). Like P11 said: ``this type of approach alone can provide a good foundation, especially in the initial stage, but soon hits the ceiling on the coverage.'' The approaches are also limited by the resources, like P9-10 questioned: ``how many resources can be assigned?'' Last but not least, experts are subjective in many senses. As explained by P2: ``experts in engineering are still not aware of everything about real-world traffic, and may be largely impacted by their experiences and how they perceive risks, which is very individual and subjective.'' 

\subsubsection{CSI approaches -- Data-driven}
\label{sec:data-driven}

Data-driven approaches are used among the interviewees and refer to identifying critical scenarios with different data collected. Like people argue that knowledge is implicitly used in any approach, data is also considered a basis for developing and testing every autonomous driving system by P12-13. One example of using this approach is that the engineering team is leveraging different data statistics to define the system requirements, including testing. Another common practice, as described by P11, is to extract scenarios from collected data sets and re-simulate them for testing, especially the hazardous scenarios, which may be rare and impractical to test in real-world traffic. Both P5 and P8 shared the same view that data, e.g., naturalistic driving data and accident reports, is a fundamental way to collect and identify more unknown and critical scenarios for testing, e.g., post-accident analysis.  

One example of a concrete application of data-driven CSI was given by P8 
regarding the testing of AEB (Autonomous Emergency Braking) system in C4. A big and representative data set (on a scale of hundreds of thousands of kilometers of driving) is collected for testing the AEB system. Then, they define the hazardous events, e.g., a false full-break, for this system and search in the data set for all triggers of such events. Lastly, they use those critical scenarios (triggers) to test the functionality and minimize hazardous events from happening again. P8 also mentioned collecting fleet test data to find unknown insufficiencies and triggering conditions, as well as acceptable levels for testing different sensors. In comparison, P6-7 shared a new perspective for smaller organizations that do not deploy vehicle fleets and sensors to collect data by themselves. They harness data and reports from third parties (e.g., authorities, companies) who provide data and analysis to the public.

The interviewees revealed five major limitations of the data-driven approaches as of today. First, the approach comes with different \emph{costs}, as described by P11, so ``we need to collect the data, annotate the data, define the criteria, and extract critical scenarios from it. In order to keep the data relevant for testing, the data also has to grow as the system evolves. Eventually, there are huge costs associated with data processing.'' Second, there are discussions over the \emph{quality} of the collected data. P2 said one preliminary problem is ``how representative the data is in terms of the coverage and distribution of different scenarios. Some data, e.g., accidents for human drivers (e.g., alcohol, distraction, speeding), may not be relevant for autonomous vehicles.'' Also, data collection is subject to the sensors and sensor setup, as indicated by P3 and P9-10. ``So, when we change the sensor setup, we might end up re-collecting data and scenarios again.'' That's why P2 articulated that besides collecting data with vehicle fleets, ``we should use infrastructure sensors that are installed on road infrastructures. The reason is that, with vehicle fleets, the collected data is dependent on the sensor setup and the driver behaviors, but with infrastructure sensors, we collect scenarios with different interactions between the road users from an independent lens.''

Thirdly, P2 presented an interesting argument about the \emph{relevance} of naturalistic driving data collected. Such data may not be representative of the autonomous driving system. On one hand, a critical scenario for a human driver, e.g., an accident due to speeding to 200 km/h or drunk driving, may not be relevant for an autonomous driving system, since they are expected to follow the speed limit and safety regulations. On the other hand, there are many scenarios where human drivers may anticipate the risks and avoid the accidents, e.g., in cases of mechanical failures or invasive behaviors of other road users, which are relevant and critical for autonomous vehicles, but never reported. Therefore, they need to address whether the critical scenarios collected are relevant for testing the autonomous driving systems, and whether we have collected all relevant critical scenarios or not. Fourthly, with all the data acquired, such as driving, road, and weather data, they need to build their competencies to \emph{understand the data} and identify critical scenarios from it, as raised by P3 and P4. Finally, P11 highlighted the current limitation of data-driven approaches in relation to the quality of \emph{simulation tools}. One significant concern is whether they have a good simulation tool to faithfully represent the scenarios as they were in the real world.

Instead of collecting substantial data from real-world traffic and extracting as well as simulating critical scenarios from it, P2 described another form of data-driven approaches -- online monitoring, which means monitoring the functioning system and recording the system operation data. It is also called the ``Shadowing function'' by P1, where the system is deployed on the vehicle but not active in controlling the vehicle. It takes all the sensory input and reacts in the background. From the system reactions, critical scenarios can be identified and used for testing and improving the system. This approach enables testing the newly developed system in its real operational environment without harming the road users and identifying more unknown and hazardous events. P2 proposed the same idea but to compare the decision from the shadowing function and the operating function, and see how they differ and find more critical scenarios by evaluating the decisions. 

\subsubsection{CSI approaches -- Search-based}
\label{sec:search-based}

In general, very few interviewees described using search-based approaches for identifying critical test scenarios, but all were positive and showed interest in this approach. As P11 said, they are not looking too much into search-based approaches, but they should invest more since more testing approaches are needed. Also, P1 said it is not used but could be effective since it is an efficient way of identifying more critical test scenarios through simulation. P9-10 claimed that this approach could generate more test scenarios and support exploring system performance boundaries. As a few examples, P12-13 used search-based approaches and critical scenarios in exploratory testing to identify system insufficiencies. P9-10 used similar approaches to manipulate and create critical scenarios (e.g., partially occluded images, rotated images) to test and improve their deep learning-based perception systems for autonomous driving.

For the time being, \textit{missing hardware aspects} are an evident limitation of this approach. As P6-7 observed based on their experiences, hardware specifications, e.g., the quality of the sensors, are usually not included in search-based critical scenario identification. Besides, P9-10 did not doubt that they could get many critical (e.g., unsafe) scenarios by searching in the parameter space. Still, the scenarios are not very useful until they figure out how to interpret the results. In other words, what do those critical scenarios mean, and what can be learned from them? In addition, the \textit{relevance} of the identified critical scenarios is a major concern commonly discussed by the interviewees. They need to understand what is tested (raised by P11) and how relevant the critical scenarios are for testing (by P2). Specifically, they have to figure out the occurrence rate or probability of the critical scenarios in the real world. ``With search-based approaches, we could identify critical scenarios that are unrealistic or have an extremely low probability of occurring'', said P6-7. P11's experience mirrors this observation, where typically, very few test scenarios are related to the perception system they own, while a large number of tested scenarios are irrelevant and unnecessary. Lastly, a good sensor simulation is needed; otherwise, the search process will be just optimizing critical scenarios for the post-perception part of the system, as described by P11.

As P5 emphasized, more research is needed on the systematic selection of relevant parameters for scenario parameterization to use search-based approaches effectively. First, \textit{realistic human behavior} (e.g., malicious behavior from other road users) should be included in the scenario exploration, as the major cause of autonomous vehicle accidents in the US as reported (P5 did not mention the specific source of it), is the misuse of autonomous vehicles by humans. Those malicious human behavior have to be identified to test the vehicles. Also, autonomous vehicles should learn to perceive and evaluate risks, understand the intentions of other road users, and improve driving capabilities over time, like humans do. They argued that ``it is unfair to just accuse humans of causing a lot of road accidents, but humans also avoided many accidents, even in critical situations. Those scenarios should also be identified and tested for autonomous vehicles as well.'' 

Second, the \textit{hardware perspectives} have to be included to generate more realistic scenarios. P11 described integrating sensor specifications in the process to identify more realistic critical scenarios for the perception system. Otherwise, they will sub-optimize critical scenarios for the post-perception part of the autonomous driving system, and fixing those critical scenarios might introduce new problems to the system. Thirdly, P3 proposed to include the system \textit{architectural dependencies} in the scenario searching process, as there is a trend towards centralized computers (i.e., one computer for all software systems on the vehicle) to save the overall cost. The dependencies among the different parts of the system or across different systems can expose new critical scenarios, and changing one part may indirectly affect other parts during operation.

Lastly, the \textit{relevance of the scenarios} have to be included in critical scenario identification. As P6-7 said: ``the probability of the occurrence of scenarios should be added to the search-based approaches. One way could be to derive the probability based on the distribution of scenarios from real-world data sets.'' P2 described exactly the same vision -- ``instead of evaluating the scenarios after, we should include the relevance in the search process in the first place.'' Nonetheless, they also point out that there will always be some residual risks -- some critical scenarios can happen with an extremely low probability, so an acceptance level has to be set (like in the aerospace domain) and restrict the maximum residual risks (based on both criticality and probability) to be tolerated. Thus, residual risks must be evaluated when searching for new critical scenarios.

\subsubsection{CSI approaches -- Combined}
\label{sec:combined}

As P11 articulated, all different approaches must be combined, as there are no silver bullets for solving critical scenario identification, and every approach has its limitations (by P8). \emph{Knowledge-based approaches}, e.g., using experts, give a short-term win and can quickly establish an initial set of critical scenarios for testing purposes, but coverage is an apparent bottleneck. \emph{Search-based approaches} enable us to explore more unknown and critical scenarios and give us better coverage, but the relevance of the identified scenarios is unsure. Likewise, \emph{data-driven approaches} are always needed since real-world data is important and essential, but the cost is a main constraint. As an example, P11 and their team used fully synthetic data, partially synthetic data, and real-world data for their simulation testing. In that process, all approaches mentioned were used. However, they did argue that the selection of the critical scenario identification approaches also depends on the scope of the ODD. Knowledge-based approaches, like experts, might be enough for systems designed for a restricted (e.g., in a more confined area) ODD, where most critical scenarios can be recognized, and safe fall-back strategies for the unknown ones could be defined. That argument was denied in another interview with P5; they claimed ``it is highly unlikely to know every possible hazardous event, especially the anomaly situation, even in a more limited ODD. For example, falling objects like flying wheels have caused many accidents every year, yet it's hard to identify every possible scenario with a falling object in testing.'' Therefore, none of the approaches can stand independently and must be combined and complemented.

P8 described that all three aforementioned approaches are used to test and find insufficiencies and triggering conditions for different systems in their autonomous vehicles. In more detail, they leveraged the domain experts, collected data sets, open data and reports, standards like ISO-26262 and SOTIF, etc., to identify test scenarios, including the hazardous ones, and analysed what was missing in each approach. P4 took a similar approach where they used domain experts, test fleet data, consumer data, etc., to derive critical scenarios. In contrast, P5 used a slightly different testing strategy, where they only collected some (limited) data from the public traffic and extrapolated more unknown and critical scenarios based on that. P1 gave a concrete example of using a combination of approaches in their testing. The experts (i.e., the accident research team) analyse every accident of their customer vehicles, select the critical scenarios, and publish an annual report for testing. They traced back to the `Moose Test'~\cite{moosewiki} case in Sweden, where flip accidents took place in an evasive manoeuvre to avoid sudden-appearing obstacles. Then, the Moose Test was designed as a standard test scenario for the vehicles. Eventually, like P12-13 said: ``every approach should be considered to some extent to find more critical scenarios and for scenario-based testing in general.''

\subsubsection{Scenario validation and evaluation}

The relevance of the critical scenarios was eventually the most discussed criteria for scenario validation and evaluation, and indicates how likely a scenario can happen in a specific ODD. P4 and P8 argued that, since the ODD determines the scope of the operational environment and limits the scenario space, any validation or evaluation of the scenario relevance should be initially based on the \textit{system ODD specification}. Consequently, ``the selection (inclusion and exclusion) of test scenarios should also be based on that'', said P6-7. Apart from ODD analysis, real-world data is another important source to use, as P2 proposed. ``We could mathematically or logically validate the relevance of the critical scenarios obtained, e.g., through guided search, based on the occurrence from \textit{collected driving data}. Besides, we could analyse and evaluate the critical scenarios with existing scenarios to project whether they are relevant'', said P4.

Unlike many other interviewees, P4 was not excessively concerned about the identified critical scenarios that might be irrelevant for testing. Especially, they claimed a need to identify and test more critical scenarios for the time being. Therefore, one should not  worry too much about having too many scenarios, but the other way around. P3's view supported this argument and proposed that, those (hypothetical or not-happening-for-now, as phrased by authors) scenarios (e.g., never seen in the current traffic), which might be deemed irrelevant, should still be tested, as they might happen in the future. ``Since our vehicles, infrastructures, technologies, etc., are evolving rapidly, and we have seen, for example, new vehicles (e.g., scooters) and accidents (e.g., airplanes landed on roads) over the years, new critical scenarios are expected. It is still the autonomous vehicles' responsibility to handle those scenarios, so we need to test how autonomous driving systems will react to that'', said P3. 

\subsubsection{Tools and platforms}
\label{sec:tools}
Every interviewee has mentioned using simulation tools for creating, visualizing, executing, and analysing test scenarios in general. Although current simulation tools are often criticized for not fully and faithfully representing scenarios as in the real world, P2 believes simulation is a rather efficient way to test and improve autonomous driving systems. At least, simulation provides a basis and approximation of how our autonomous vehicles perform in different scenarios and where they could fail. As said by P8, a major part of their testing for every software or software change, is to re-create and test scenarios from the collected driving data in simulation. However, quality has been considered a limitation of the existing simulation tools. As P3 stressed, there are still gaps between simulation testing and real-world testing, e.g., noises for the sensors, and there is a lack of evaluation frameworks to assess how accurate the simulation is to the real world. Thus, improvements to the simulation tools and gap analysis are needed. 

In terms of testing in relation to critical scenario identification, P3 talked about a platform that was used by some industry practitioners for identifying and testing critical scenarios for autonomous driving systems. The general idea is to load the autonomous driving systems into a racing game engine (e.g., GTA), create very challenging scenarios by adapting the dynamics (e.g., speed, acceleration, and even abnormal behaviors) for different road users, and see whether the systems can handle those scenarios as well as analyzing when the system fails.  

\subsubsection{General initiatives for testing with CSI}
\label{sec:initiatives}

Generally, the interviewees agree that it is important to use different approaches for critical scenario identification and combine them for testing. Like P9-10 said: ``expert knowledge and analysis must always be included, whether we use data-driven or search-based approaches. We need domain expertise to define what is critical and whether a critical scenario is relevant or not. Also, it is important that we continuously learn the field in terms of new tools, data, test processes, techniques, and approaches, etc.'' As P5 articulated, there are still many challenges in different areas, but they will know more and build the competencies over time. Consequently, they will iterate and evolve the technologies, methods, and practices to fill in the gaps. The point is, when they see the problems, they try to improve the situation. 

P8 proposed a similar initiative, arguing for continuous monitoring of the hazards, insufficiencies, and misbehaviors after the release of the autonomous driving functions, so that the systems will be improved, and become safer and more robust in the long run. Indeed, like P12-13 described, critical scenarios must grow and be continuously developed since new traffic infrastructure, regulations, vehicles, and behaviors appear. It can arguably be a fact that every critical scenario never will be thought of beforehand, but the system will evolve as well. Thus, ``we need to have a more continuous approach and iterate the testing, the critical scenarios, and the critical scenario identification approaches with more data, expertise, tools, etc.'', said P5. During this process, it is extremely important that different stakeholders (e.g., academia, industry, authorities, and the general public) collaborate instead of working on their own. A typical example, as described by P9, is to build a data eco-system so that every organization can share their data and technologies for processing the data. That is an initiative where they still think academia could play an important role since they are more neutral and involves no direct competition or conflicts of interest with the manufacturers.   

Remarkably, there is another initiative proposed by several interviewees. P5 suggested including a fail-operational mechanism, allowing the autonomous driving systems to report unknown critical scenarios and request interventions, e.g., remote control. In other words, the risk have to tolerate that some critical scenarios, that are unknown and not tested, will occur in the real world. Like P9-10 said, there could be some scenarios which are unavoidable collisions, so the system can only mitigate and minimize the damage as much as possible. As a potential solution, P11 claimed and stressed that ``we should not over-focus on finding all critical scenarios, but improve the system to handle in case unknown scenarios take place''. To that end, ``we need to set an acceptable level for the residual risks -- some scenarios may have a high severity but an extremely low probability'', said P12-13. Also, reasonable expectations need to be set for autonomous driving systems -- not identifying and resolving all collision and hazardous scenarios, but minimizing the potential consequences through, e.g., defensive driving. A recent example from social media, as given by P11, is that a Tesla autonomous vehicle encountered a horse carriage on the road and was not able to classify it. It would be reasonable not to have such a scenario in the test, but fall-backs e.g., slow-down or pull-over the vehicle, so that the vehicle will not hit it and cause any accidents.

\subsection{Challenges of testing with CSI (RQ2)}
\label{sec:challenges}

\begin{figure}[t]
\includegraphics[width=\textwidth]{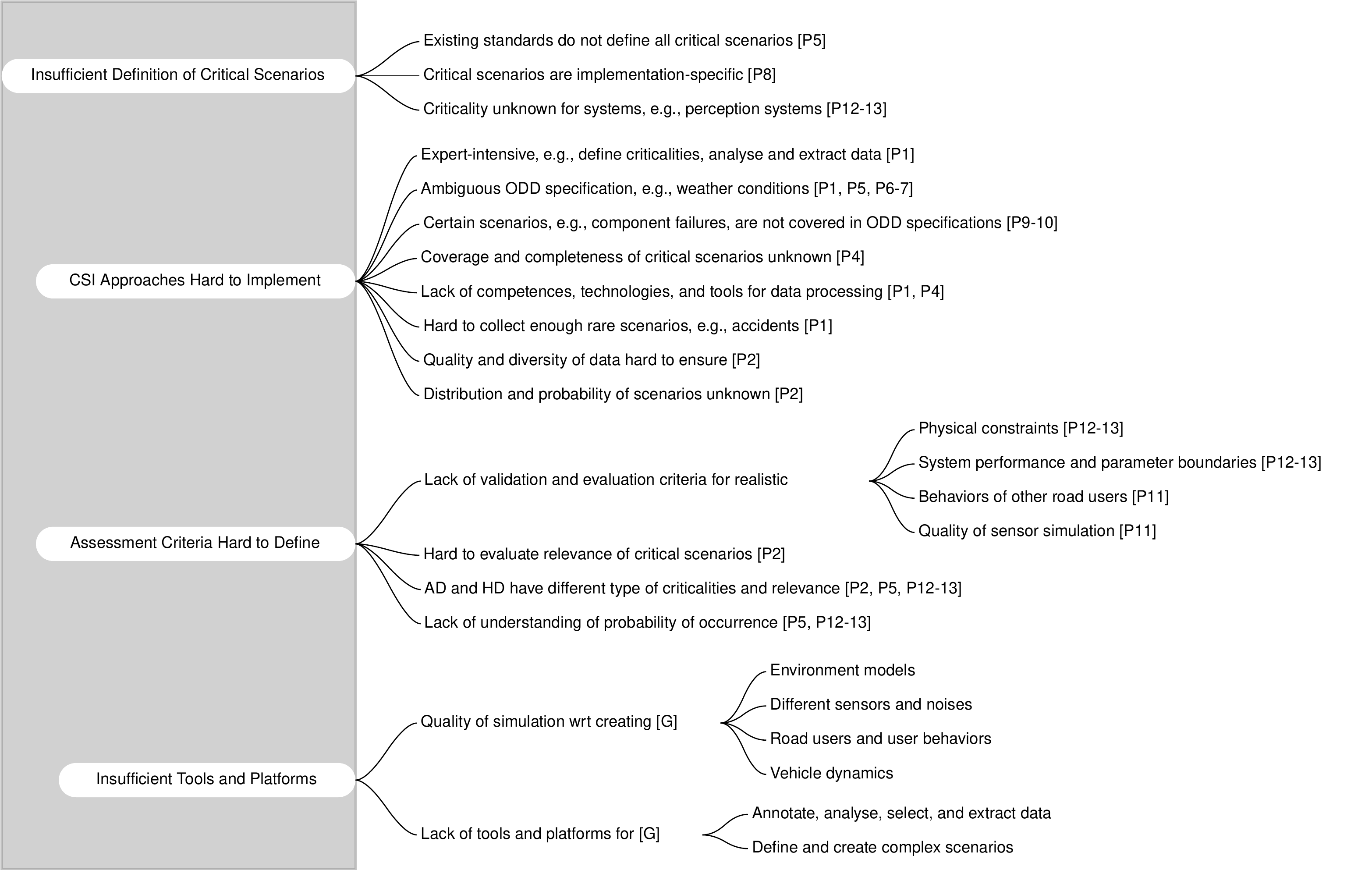}
\caption{A summary thematic model of (RQ2) challenges for critical scenario identification and testing. [G] represents a general challenge that is commonly raised by all interviewees. AD is short for autonomous driving, and HD is for human driving.}
\label{fig: challenges}
\end{figure}

In this section, we mainly focus on presenting the challenges of testing autonomous driving systems with critical scenario identification. Similar to Section~\ref{sec:practices}, we primarily present challenges in relation to critical scenarios and critical scenario identification approaches. A summary synthesis model for the challenges is presented in Fig.~\ref{fig: challenges}, and the (four) top-level themes in the grey rectangle are further described and elaborated in Sections~\ref{sec:insufficient_definition}--\ref{sec:insufficient_tools} respectively, in top-down order. The model is summarized from the full thematic model of the study, and includes the four top-level themes under RQ2 and the sub-themes on
the appropriate level of abstraction. The sub-themes are embodied with minimum explanation and context from the codes and interview transcripts to be self-explanatory.

\subsubsection{Insufficient definition of critical scenarios}
\label{sec:insufficient_definition}

Current standards and regulations do not explicitly define and list all possible critical scenarios; thus, the definition is considered too general. As P5 said, the SOTIF standard does provide some examples of critical scenarios, such as misuse or abuse scenarios. Still, many identified critical scenarios are not included yet, like critical scenarios of hijacking an autonomous vehicle. In addition, critical scenarios are \textit{implementation-specific}, which means that critical scenarios are different for different implementations of the system. P8 discussed that every company has its own strategies for avoiding hazards and using different sensors. For example, Tesla uses vision (cameras) only, but others may also combine radars and lidars. As a result, every organization invents their own definition based on their understanding of the standards and system implementations. Given that, critical scenarios from one organization may not apply to another, even for the systems with the same functionalities.

Generally saying, it is very difficult to define which scenarios are critical. As P12-13 described: ``we don't really understand what the critical scenarios are or could be for the perception system. Even though we can explore more unknown scenarios, and try to solve the problematic ones (e.g., minimizing the accident rate, mitigating the potential damage and consequences) we found, we can not just rely on the extreme cases and assume other scenarios would be handled correctly. That is particularly true for the (deep learning-based) perception systems, where it is hard to quantify and estimate the criticality and say one scenario is more critical than others.''

\subsubsection{CSI approaches hard to implement}

As P1 articulated: ``critical scenario identification and testing are \textit{expert-intensive}, regardless of which approaches (e.g., data-driven, search-based) we use.'' Expert knowledge is needed to define, derive, and identify critical scenarios; otherwise, it is hard to know what is critical, how realistic a critical scenario is, how good the simulation is, etc. In that sense, the outcome is heavily relying on expert knowledge to determine, including but not limited to, what data to collect, how to analyse them, and which data are more critical. Also, a primary challenge with critical scenario identification, as described by P1 and P6-7, is clearly defining the \textit{ODD specification}, which is often ambiguous for nuanced parameter selection for search-based approaches. For example, how to quantify the weather, such as rain, fog, and snow, and what a dry road condition refers to. ``Without a full and explicit ODD specification, it is very hard to select all relevant parameters in a more comprehensive and systematic way'', as said by P5.

Critical scenario-based testing, or even scenario-based testing in general, can only cover parameters in the ODD specification, such as weather, speed limit, and road curvature, and nominal scenarios, such as lane-keeping or collision avoidance. However, there are still many parameters and critical scenarios, e.g., fault injection, not covered in the ODD specifications. ``An example is the system or component failure, which is not specified in the ODD, and is subsequently not covered in the scenario-based testing'', said P9-10. ``The problem is we can not cover all relevant parameters when using the critical scenario identification approaches; thus, there would always be some critical scenarios missing and not tested'', as said by P4. As a result, the \textit{coverage and completeness} of the critical scenarios are unknown and questionable.

The challenge in implementing the critical scenario identification approaches also owes to the fact that there is a lack of technologies (e.g., techniques, tools, approaches, etc.) and competence for \textit{data processing}. P1 described that a problem with data collection is that substantial data is needed to train machine learning models. Still, some data samples, e.g., accidents, are rare and can only be collected with a large vehicle fleet and immense driving mileage, resulting in enormous costs. Another concern is the diversity of the data, where a large data set has to be collected in order to cover different scenarios (raised by P2). Further, the distribution and probability of the scenarios have to be identified to use them in critical scenario identification; otherwise, irrelevant or even unrealistic critical scenarios might be derived or identified. After all, with the size and complexity of today's data collection, they need to build our competence and technologies to understand and use the data, as P4 described. ``We still face many challenges in selecting data, setting the criteria, identifying and evaluating the critical scenarios, etc.'' That is yet another challenge that P4 thought academic researchers could step into and support the industry side.

\subsubsection{Assessment criteria hard to define}

Assessment of the critical scenarios was another challenge commonly raised by many interviewees. Firstly, there is a lack of criteria to evaluate the \textit{realism} of the critical scenarios. P12-13 articulated that it is important to figure out all the constraints (e.g., physical) and the boundaries (e.g., speed limit, road curvature, and inclination range), and a combination of them as some constraints may relate to or depend on the others, to understand the realism of the critical scenarios. Critical scenarios from knowledge-based or search-based approaches can be unrealistic if physical constraints, such as system reaction time, break activation, tire friction, under a certain speed and road curvature, are not considered. In other words, they need to evaluate whether some maneuvers are feasible to know whether the scenarios are realistic. Also, P11 highlighted that critical scenarios could be unrealistic due to the behaviors of other vehicles or road users involved in the same scenario, e.g., how a vehicle performs cut-in, or the simulation is unrealistic, e.g., the lidar simulation gives sensory data that is different from the actual radar for the perception system. Nevertheless, it is hard to define effective criteria to assess the realism of the critical scenarios from, for example, search-based approaches, since the resulting critical scenarios have to be interpreted and evaluated in a reliable way. However, that view was denied by P9-10, who thought the critical scenarios are unrealistic because they do not strictly or precisely define the searching constraints and targets, e.g., limiting the relative speed or distance; otherwise, the search-based approaches should identify realistic critical scenarios that can happen in the real-world traffic.  

Secondly, the \textit{relevance} of the critical scenarios can hardly be evaluated as of today. Relevance, different from realism, measures how applicable a scenario is to a certain system and the system functionalities we are testing. As P2 described, identifying a critical scenario, e.g., searching in simulation, may not be too difficult, but understanding its relevance is tricky, even for a simple scenario like a vehicle behind performs a cut-in for a lane-keeping function. P2 referred to an international conference on automotive and autonomous driving, where participants agreed on the significance of relevance but had no answers on how to solve it. They also stressed the need to create a comprehensive scenario catalog and study the relevance of the scenarios in a specific ODD. Similarly, P12-13 said: ``human driving has different types of criticality and relevance than autonomous driving. Critical scenarios for humans may not be as critical for autonomous vehicles, and vice versa. For example, human driving accidents due to distraction, fatigue, and drunk driving, are irrelevant to autonomous driving and should not be included in testing.'' In contrast, as P5 described, the major cause of accidents for autonomous vehicles (as reported in a certain country) is the misuse of them by humans, which is relevant, and should be identified and used for testing.   

Thirdly, there is still a lack of understanding of the \emph{probability} of the critical scenarios identified. Like how P5 and P12-13 stressed, a challenge with critical scenarios is that they need to understand the likelihood of the occurrence in real-world traffic. P5 also mentioned the frequency, which indicates how often a critical scenario has happened in real traffic. While testing critical scenarios is mandatory, the problem is whether they have tested enough, if not all, critical scenarios. They wouldn't know all possible critical scenarios, even for a low-level autonomous driving function, especially anomaly situations, e.g., serial crashes or falling objects. A critical scenario may only be added to the testing after it has happened. An alternative is to identify and cover more unknown scenarios, as much as possible. Nevertheless, it is hard to demonstrate the completeness of the testing. As P8 said: ``we can not define and predict all unknown scenarios since they are unknown, so we can only test the robustness of the system for any surprising inputs, e.g., unknown hazardous scenarios, that the vehicle can still manage to some extent.'' P5 shared the same view that ``there will be unknown scenarios out there, and there are still technical gaps in recognizing all the objects and events, like different falling objects or animals. Still, we need to make sure our systems can detect and avoid such objects even without knowing what they are.'' 


\subsubsection{Insufficient tools and platforms}
\label{sec:insufficient_tools}

As has been mentioned in different places (Sections \ref{sec:data-driven}, \ref{sec:search-based}, and \ref{sec:tools}), a challenge associated with tools and platforms for critical scenario identification and testing is the quality of the simulation tools. As P11 articulated, the biggest challenge for executing, extracting, or searching for critical scenarios, is to have a good simulation of the environment, sensors, and vehicle dynamics etc., so the simulation tool can provide a faithful representation of real-world traffic. Currently, the creation of such an environment and models for how different parts (e.g., sensors, pedestrian behaviors) work is still manual, and the quality is clearly insufficient for a realistic environment. The potential risk is identifying critical scenarios that are unrealistic or irrelevant. Also, it is generally acknowledged by the interviewees that more tools and platforms are needed to annotate, analyse, and extract data, as well as to create complex scenarios.

\section{Discussion}
\label{sec:discussions}

\begin{figure}[t]
\includegraphics[width=\textwidth]{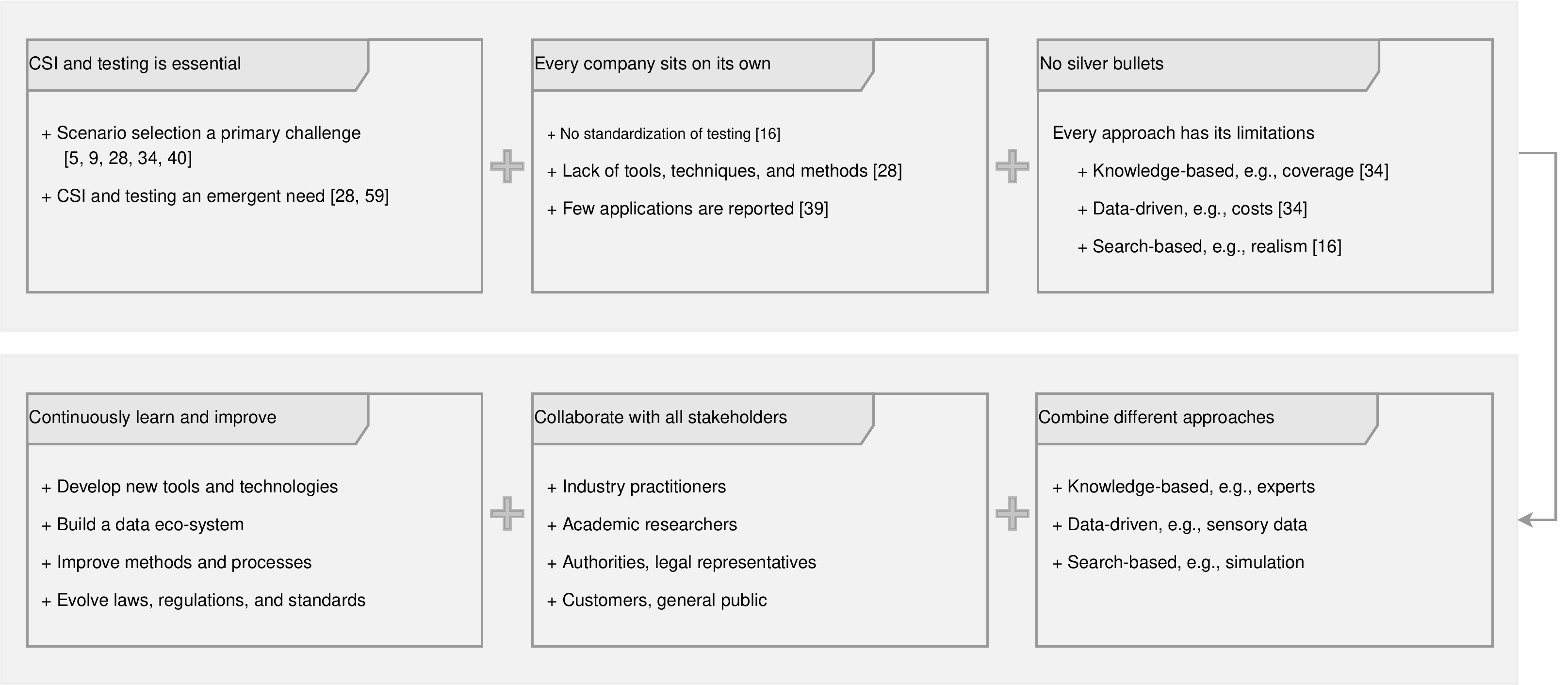}
\caption{A summary of the findings of this study. The three cards in the upper and grey rectangle summarize findings of the background, practices (RQ1), and challenges (RQ2) of critical scenario identification and testing, respectively. The three cards in the lower and grey rectangle contain the three recommendations (combine, collaborate, and continuously learn) synthesized based the interviews and our analysis.}
\label{fig: conceptual_model}
\end{figure}

In this section, we elaborate on the findings of the study in conjunction with the relevant literature (i.e., in Section~\ref{sec:discussion_limitations}), as shown in Fig~\ref{fig: conceptual_model}. We do not repeat the practices, as presented in Section~\ref{sec:results}, but discuss the views and insights based on that and the relevant literature studies. Also, we provide our own analysis and insights on what we could do next (i.e., in Section~\ref{sec:discussion_combine}) regarding the current limitations and challenges. In Section~\ref{sec:threats}, we particularly discuss the validity of the study and how we have addressed or mitigated the potential threats to validity.

\subsection{No single universal CSI approach}
\label{sec:discussion_limitations}

In general, the scenario-based approach is widely used for testing autonomous driving systems, yet scenario selection and elicitation is a primary challenge~\cite{riedmaier2020survey, lou2022testing, neurohr2020fundamental, beringhoff2022thirty, de2020procedure} to solve. That is, which scenarios to test, where to get them, and how to demonstrate the testing with the selected scenarios~\cite{beringhoff2022thirty}. Further, it also concerns the test execution and evaluation, and how to use the test results for safety argumentation~\cite{neurohr2020fundamental}. Critical scenario identification is a part of scenario-based testing and aims to find and test the critical scenarios for autonomous driving systems. Lou et al. has confirmed critical scenario identification as an emergent need in the industry~\cite{lou2022testing}. Just like Zhang et al. articulate, we need to understand which scenarios are critical, how to derive them, and how to use them for safety analysis~\cite{zhang2022finding}. Although the interviewees showed different interests and practices for critical scenarios, identifying and testing those scenarios is significant and essential for autonomous driving systems.  

As discussed in Sections~\ref{sec:knowledge-based} and \ref{sec:combined}, the \emph{knowledge-based} approaches serve as a good start, where the domain ontology provides a common understanding of the field and a basis for critical scenario creation~\cite{li2020ontology}. Such domain-specific ontology can be designed from expert experiences, standards, laws, regulations, etc.~\cite{lou2022testing}. Nevertheless, knowledge-based approaches may result in an incomplete set of critical scenarios~\cite{neurohr2020fundamental}, and the scenarios may lack a real-world representation~\cite{lou2022testing}. In general, as discussed by Bach et al., identifying and selecting scenarios manually is time-consuming, and relies heavily on the knowledge of the scenario designer and their interpretation of the test specifications. In addition, some rare critical scenarios can be missed, and the scenarios must be validated to ensure an acceptable level of realism before testing~\cite{bach2017test}. 

The \emph{data-driven} approaches are vital and common for critical scenario identification, as presented in Section~\ref{sec:data-driven}. The data can be in different forms and refer to the sensory data from naturalistic driving collected by vehicle fleets or infrastructure sensors, map data, accident reports, industry standards, laws, and regulations etc.~\cite{cai2022survey}. According to a survey study by Lou et al., naturalistic driving data is the most commonly used source for test scenarios~\cite{lou2022testing}. While scenarios from naturalistic driving data are said to be naturally realistic~\cite{cai2022survey}, a sufficiently large and diverse data set is the prerequisite~\cite{neurohr2020fundamental}. 
Artificial driving data are often created and used as complementary data sources, yet Olleja et al. found the criticality of scenarios from artificial data is very different from realistic data and claimed that using artificial data for testing must be extremely cautiously~\cite{olleja2022can}. Statistics from NHTSA~\cite{singh2015critical}, and Kalra and Paddock~\cite{kalra2016driving} also show that human errors have caused more than 90\% of road accidents; thus, using the accident data from human driving must be further analysed to ensure the relevance to autonomous driving. Moreover, tool support is lacking and rated as the most urgent need for data-driven approaches in Lou et al., where we need effective tools for, e.g., labelling data (tagging bounding boxes, semantic segmentation masks), defining complex scenarios~\cite{lou2022testing}. 

The \emph{search-based} approaches use search algorithms to optimize critical scenarios with respect to the objective functions for quantifying the criticality~\cite{riedmaier2020survey, lou2022testing, kolb2021fitness}. This approach has received increasing interest in academic research, yet has not been implemented equally in the industry, as our interviewees describe in Section~\ref{sec:search-based}. A major challenge is to incorporate realistic sensor performance~\cite{beringhoff2022thirty} and behavior models for other road users, e.g., human drivers, pedestrians, cyclists, animals, etc., to identify more unknown and realistic critical scenarios~\cite{hacohen2022autonomous}. Petrovic et al. and Favaro et al. studied the accidents of autonomous vehicles in California, US, and both found that the most common type of accident was rear-end collisions where an autonomous vehicle was hit by a human-driving vehicle and inferred that human drivers were more aggressive and not accustomed to the driving style of autonomous vehicles in traffic~\cite{petrovic2020traffic, favaro2017examining}. Realistic behavior models of possible road users must be included in the critical scenario identification, and so does the systematic parameter selection~\cite{scholtes20216}. 

Another prominent challenge is to select good \emph{objective functions} (also referred to as fitness functions) for implementing search-based approaches. Even though there are various criticality metrics~\cite{westhofen2023criticality} and safety performance evaluation metrics~\cite{wishart2020driving} available for autonomous driving, no single metric can capture all critical aspects of all scenarios~\cite{neurohr2020fundamental}. Just like Cai et al. discussed, TTC might be enough for car-following scenarios, but not others~\cite{cai2022survey}. Thus, objective functions must be selected attentively~\cite{kolb2021fitness}, and multiple objective functions may be combined~\cite{cai2022survey}; otherwise, safety can hardly be demonstrated. Simply selecting an objective function can be problematic, for example, minimising the distance between vehicles may identify critical scenarios with short distances in between, but low distance alone cannot demonstrate the criticality without considering the speed~\cite{kolb2021fitness}. When using search-based approaches, it is important to ensure the reasoning of selection or derivation of the objection functions, or the test results may not provide the expected confidence for the safety argument~\cite{kolb2021fitness}. 

Also, the \emph{gap between the real world and simulation environments} is a clear limitation~\cite{beringhoff2022thirty}, and some critical scenarios (with adversarial samples) which might be easily identified in the simulation are not correlated with real-world scenarios, thus are not transferrable to the real-world testing~\cite{stocco2022mind}. Like Adigun et al. explained, ML-based components can be easily misled by unexpected (adversarial) inputs, and deriving conclusions from such testing results is intricate, as they are often unrealistic since they do not consider the full operational environment and its semantics~\cite{adigun2022collaborative}. Kolb et al. argued that not all possible scenarios are necessarily good test scenarios as they may be irrelevant (e.g., not containing the manuever under test) or not challenging enough (e.g., having a far distance between vehicles)~\cite{kolb2021fitness}. As examples, Sun et al. excluded critical scenarios with unavoidable collisions in the initial state as they claimed those scenarios are irrational and meaningless for testing~\cite{sun2021scenario}; Kluck et al. separated scenarios with TTC larger than 20 seconds as invalid for testing an autonomous emergency braking system~\cite{kluck2019genetic}. Even though it is debatable whether their arguments are appropriate or not, some validation and evaluation criteria for the realistic critical scenarios are needed anyway. Lastly, as described by Cai et al., no critical scenarios found in the optimization and simulation do not mean the system is safe~\cite{cai2022survey}. Generally speaking, as said by interviewee P11, the search-based approaches are considered effective for critical scenario identification and should be invested more in the future.

Overall, every approach has its own strengths and weaknesses, and thus none of them can stand on its own. In general, the knowledge-based approaches provide a baseline but are subject to the coverage; data-driven approaches are common and useful, yet rely on a large and diverse data set; search-based approaches can identify more unknown critical scenarios effectively, while the realism and relevance of the resulting critical scenarios need to be further analyzed. Besides, there are several common limitations to some or all of these approaches. First, as Cai et al. claimed, even though scenario-based approaches are widely used (in different studies) for testing autonomous vehicles, no one concludes \emph{how many scenarios would be enough}~\cite{cai2022survey}. Second, the \emph{dynamic feasibility} (due to physical constraints) of given maneuvers and the composition of different maneuvers must be analysed to explore realistic critical scenarios~\cite{Abbas2017driver}. Thirdly, the severity of critical scenarios must be accompanied by the \emph{probability} of their exposures, as risk is typically dependent on both the exposure and criticality~\cite{tenbrock2021conscend}, so arguing criticality alone without knowing the exposure is in-comprehensive. Lastly, we need to understand how to \emph{validate the simulation tool and environment}~\cite{neurohr2020fundamental}, like Yu et al. articulated, a good simulation is a prerequisite for using digital twins for testing autonomous driving~\cite{yu2022autonomous}. Recent EU regulation has mandated evaluating the credibility of modeling and simulation as part of the type-approval for autonomous vehicles~\cite{eu2022}.  

\subsection{Combine CSI approaches and collaborate}
\label{sec:discussion_combine}

As a general initiative proposed by our interviewees (i.e., Section~\ref{sec:initiatives}), all approaches should be considered and \emph{combined} for critical scenario identification. In particular, P8 set a very good example (as discussed in Section~\ref{sec:combined}), where they used all approaches for identifying critical scenarios and testing their autonomous driving systems. They also analysed what is missing in each approach and how one could complement the others. As Cai et al. stressed, using data-driven approaches does not necessarily mean expert knowledge is unnecessary. Instead, expert knowledge is utilized in all data-driven approaches or at least as complementary to them~\cite{cai2022survey}. Furthermore, Neurohr et al. discussed that knowledge-based approaches, e.g., using experts, and data-driven approaches can be combined and used either in a top-down manner, for example, experts derive relevant scenarios and then use data for analysing the distribution or probability, or in a bottom-up manner, for example, extract scenarios from data, and then use experts to classify the scenarios~\cite{neurohr2020fundamental}. Also, instead of having every organization sit on its own and creating everything by itself, there is a strong call from our interviewees for immediate \emph{collaboration} among different entities, such as the automotive industry, academia, legal authorities, user communities, and even the general public, to contribute together in evolving the tools, technologies, methods, processes, data, and regulations, etc. for testing and improving the autonomous driving systems. Indeed, as P8 said: ``we need to build an eco-system of everything we need, and with that, we could be more efficient and enable more people or organizations to be involved.'' \emph{Continuous learning} is also an important aspect raised by our interviewees, and ``we need to keep evolving different perspectives of the field and improve the technologies over time. We build things step by step'', as articulated by P5.

As articulated by Reisgys et al., future testing methods for autonomous driving should aim to satisfy requirements such as more in simulation, increased coverage of test scenarios, increased test validity (e.g., the validity of simulation and scenarios), minimizing unknown scenarios, and efficient scenario prioritization based on relevance~\cite{reisgys2022scenario}. Eventually, we have to combine all resources and approaches for that. Also, as discussed by Cai et al., both scenario coverage and critical scenarios are essential for testing autonomous driving systems, because coverage ensures the autonomous driving systems for different scenarios, and critical scenarios can quickly find more faults of the systems~\cite{cai2022survey}. Neurohr et al. claimed that since new scenarios may emerge in the operational environment of autonomous vehicles, we need to develop an update process and constantly monitor autonomous vehicles during their operations to identify more unknown and critical scenarios~\cite{neurohr2020fundamental}. Despite all the challenges we discussed in identifying and testing critical scenarios, there are some relevant studies that shed light on them, which we would like to highlight. For example, Bach et al. proposed a scenario database and structured framework for selecting relevant test scenarios based on the geo-location characteristics of them~\cite{bach2017test}. Birchler et al. claimed previous studies on testing autonomous vehicles using simulations have shown that most of the auto-generated test scenarios do not significantly improve the quality and reliability of the systems, and those scenarios are considered uninformative and should be selected to avoid wasting test resources. To solve that issue, they developed an ML-based approach for selecting critical test scenarios that expose faults before actually executing them in the test~\cite{Birchler2022cost}. Yu et al. proposed using kinematic model-based intelligent agents to control the traffic states and avoid unrealistic interactions between different road users, e.g., a vehicle accelerates and hits an autonomous vehicle in front.~\cite{yu2022autonomous} Nakamura et al. proposed to parameterize scenarios from real-world driving data, learn their distribution, and define the reasonably foreseeable parameter ranges based on that~\cite{Nakamura2022defining}.

\subsection{Threats to validity}
\label{sec:threats}
We adopt the framework described in Runeson et al.~\cite{runeson2009guidelines} to evaluate the validity of our study and discuss how we have mitigated potential threats to the validity. Specifically, we present the validity from the internal, external, and construct perspectives, and we particularly focus on internal and construct validity as the study is exploratory and qualitative.

Internal validity refers to the validity of the results internal to the study, and concerns such as how did we collect data, analyse data, and draw conclusions based on that. Initially, we used a convenient sampling approach to get relevant interview candidates from our network, based on their experience and expertise in the related field. Afterwards, we mainly used a snowballing approach, where the interviewees recommended us more candidates that they knew and who worked with the topic we study. Therefore, we ensured that the participants we interviewed were experts in this domain and knew the topic well. Also, the interviewees had different roles and worked on various projects as shown in Table~\ref{table:interviewees}, and they covered a wide spectrum of companies (i.e., seven) that we knew in the autonomous driving domain in Sweden. 

Furthermore, we reviewed and kept our interview questions simple, clear, and open to the interviewees. We explained and elaborated on them to ensure the interviewees understood and were comfortable answering our questions. In addition, we sent the transcription back to the interviewees to make sure every part of the interview was clear to us, and they could adapt their answers if they wanted. Given that, we tried to maximize the validity of the data with respect to the relevance to our questions and the factual accuracy of what they said or meant. Regarding the analysis and synthesization of the interview data, we coded and modeled the data in multiple steps to improve the accuracy of our interpretation, and we used cross-review among the authors to maintain the consistency of our analysis. As discussed in Section~\ref{sec:thematic_analysis}, we had several seminars to discuss the synthesis, resolve the disagreements, and keep us, as well as the thematic modeling more aligned and coherent.

External validity refers to the generalizability of the results to other, or entire, populations in the field, or even across fields. In our study, we aimed to explore the industry practices to understand how critical scenario identification has been used for testing and related challenges that exist. However, we do not claim that the results are comprehensive or consistent with other practitioners in this domain. In other words, it is possible that the views or insights we present in this study are denied by other practitioners, and they may propose new practices and challenges based on their experience and focus. That said, as we have explained above, we interviewed 13 experts from seven different companies in Sweden which are at the forefront of autonomous driving, thus we believe the findings are representative to some extent. The external validity is strengthened by the fact that the participants come from different companies, teams, roles, and experiences. To the extent of our knowledge, the companies we included in the study cover the main players in the autonomous driving domain in Sweden, and are connected to other international companies (e.g., in China and Germany) in this field through, e.g., collaboration or shared ownership.

Construct validity concerns whether we are studying the right things with respect to the research questions and objectives in relation to the topic of the study. We have studied the literature in the field, discussed the research gaps, and proposed two explicit questions in this study, as described and listed in Section~\ref{sec:introduction}. 
We selected interviews as the data collection method in this study because exploring industry practices entails we have many open questions~\cite{linaker2015survey}, where context information is required to understand the questions, and follow-up discussions are highly expected to extend the answers. Before starting each interview, we presented an overview of related terms, concepts, and studies for such as scenarios, critical scenarios, and critical scenario identification, as described in Section~\ref{sec:related_work_terms}, to the interviewees, to set up the common ground for the interview. 

During the study, we carefully defined the interview questions, reviewed them together, and explained them when necessary, to ensure that we got valid answers relevant to the topic of the study. We did not make any hypotheses or set any expectations with respect to our research questions, or interviewees' responses. We kept the questions open and flexible, and we aimed to understand and expand their answers instead of judging the responses in any form. Besides, we conducted a pilot interview to testify that the interview protocol and interview questions were properly designed. We had at least two interviewers in each interview to ensure the interviews were on the right track and avoid missing anything important. Before the interview started, we always introduced our study, terminologies, concepts, different approaches, and example studies of the approaches in the area to our interviewees, to set up the common ground for the interview.

\section{Conclusions}
\label{sec:conclusions}

Scenario-based testing is a common approach for testing autonomous driving systems, and a primary challenge is to derive and select relevant test scenarios, especially the critical ones that can cause hazards or risks of harm to the autonomous vehicle and other road users. Even though different approaches have been studied for identifying critical scenarios, there are very few applications of such approaches reported from the industry, and consequently, the practices and challenges are not well understood. To address that, and learn about practices (RQ1) and challenges (RQ2) of using critical scenarios for testing, we conducted 10 interviews with 13 practitioners from 7 companies in the autonomous driving domain in Sweden. 

For RQ1, it is generally accepted by our interviewees that critical scenario identification and testing are essential for assuring the safety and reliability of autonomous driving systems. However, every company sits on its own today, exploring different approaches and tools. There is no standardization for identifying and using critical scenarios for testing its autonomous driving systems. As for RQ2, it is clear that there are no silver bullets for critical scenario identification and testing, as every approach has its weaknesses. Thus, we recommend, based on the interviews and analysis, different approaches available should be \emph{combined}. Also, different stakeholders in autonomous driving domain must \emph{collaborate}, and \emph{continuously learn} the field with respect to new tools, techniques, methods, data, etc., to improve critical scenario identification and testing. 

The contribution of this study is two-fold. First, we provide insights and analysis into the current practices of critical scenario identification and testing in the industry. Second, we believe the findings give a better understanding of the actual challenges and will benefit the industry relevance of future research. Nevertheless, we notice that gaps still exist, as there is no conclusion yet on, e.g., how the existing approaches should be combined for testing, how to enable efficient collaboration, etc. Those are important questions to be considered and addressed in future work.

\section*{Acknowledgement}
This work was supported in part by the Wallenberg AI, Autonomous Systems and Software Program (WASP).

\bibliographystyle{ACM-Reference-Format}
\bibliography{reference}

\end{document}